\documentclass[superscriptaddress,preprint,12pt]{revtex4-1}
\usepackage[table]{xcolor}
\usepackage{rotating}
\usepackage{amsmath}
\usepackage{bm}
\usepackage{array}
\usepackage{graphicx}
\usepackage{dcolumn}   
\usepackage{amssymb}
\usepackage[english]{babel}

\begin{document}

\title{Local drag of a slender rod parallel to a plane wall in a viscous fluid}

\author{Lyndon Koens\footnote{lyndon.koens@mq.edu.au}}
\affiliation{ Department of Mathematics and Statistics, Macquarie University,12 Wally’s Walk, Sydney, New South Wales 2109, Australia}
\author{Thomas D. Montenegro-Johnson}
\affiliation{ School  of  Mathematics,  University  of  Birmingham,  Edgbaston,  Birmingham  B15  2TT,  UK.}
\begin{abstract}
The viscous drag on a slender rod by a wall { is important to many biological and industrial systems. This drag} critically depends on the separation between the rod and the wall { and can be approximated asymptotically in specific regimes, namely far from, or very close to, the wall, but is typically determined numerically for general separations.}  In this note we determine an asymptotic representation of the local drag for a slender rod parallel to a wall which is valid for all separations. This is possible through matching the behaviour of a rod close to the wall and a rod far from the wall. We show that the leading order drag in both these regimes { has been known since 1981} and that they can used to produce a composite representation of the drag which is valid for all separations. { This is in contrast to a sphere above a wall, where no simple uniformly valid representation exists.} We estimate the error on this composite representation as the separation increases, discuss how the results could be used as resistive-force theory { and demonstrate their use on a two-hinged swimmer above a wall.}
\end{abstract}

\maketitle

\section{Introduction}

Viscous flows around slender objects by walls occur in many important microscopic fluid systems. For example bacteria and spermatozoa swim towards boundaries using filaments called flagella \cite{Walker2019, SMITH2009a, Das2018, Spagnolie2012, Bianchi2017, Ishimoto2014}, beating hairs called cilia line our airways and help keep them clean \cite{M.Vanaki2020, Cicuta2020}, artificial microscopic machines often use fibres and need to navigate tight conditions \cite{ Lippera2020, Huang2019} and fibre-reinforced plastic machine parts can be created by injection moulding \cite{Sanjay2018, Yamanoi2010}. Though such geometries are common, the flow around slender bodies can often be tricky to model \cite{Reis2018}. This is because the large aspect ratios (length/thickness) of these objects causes numerical simulations to require a high resolution to accurately capture the flow \cite{Walker2019, Das2018, Yamanoi2010, SMITH2009a}. Hence asymptotic techniques, called slender-body theories (SBTs), are often used to simulate these systems \cite{Cox, 1976, Batchelor2006, Keller1976a, Johnson1979, Koens2018}. 

Slender-body theories expand the viscous flow around a slender object in terms of the inverse of the aspect ratio of the body. In this limit, filaments in isolation display two regions of behaviour: an inner region in which the body behaves like an infinite cylinder and an outer region in which the body is effectively a line \cite{1976}. These regions can be matched together to solve for the flow and the drag per unit length along the object. This drag per unit length is determined though an integral equation over the centreline of the body for algebraically accurate SBTs \cite{1976, Keller1976a, Johnson1979, Koens2018} or a set of local drag coefficients for logarithmically accurate ones \cite{Cox, Batchelor2006}. Though less accurate, these local drag coefficient models - often called resistive-force theories - capture the leading physics and are easy to use. Viscous SBTs have been successfully used for models of isolated micro-organisms \cite{Kim2004, Koens2014, Myerscough1989, Yang2011}, dilute suspensions of rods \cite{Tornberg2006,Nazockdast2017a}, and the dynamics of secluded elastic fibres \cite{Chakrabarti2019b, Clarke2006} and have been extended to filaments with non-circular cross-sections \cite{Batchelor2006, Koens2016, Borker2019}.   

The presence of walls complicates the slender body asymptotic expansions by introducing additional length scales. { Similarly to the viscous models for spheres by walls \cite{Kim2005, Jeffrey1984a}}, SBTs with walls are typically restricted to distinct regimes and configurations \cite{Barta1988,Brennen1977a,Yang1983, Lisicki2016b}. For example there have been several studies of rods exactly half way between parallel plane walls \cite{Takaisi1956, Katz1975, DeMestre1973} but, to the author's knowledge, none at general separations \cite{Liron1976}. This issue is also present in the simple case of rod by a single plane wall. In this case there exists asymptotic solutions in the limit that the separation is much larger then all lengths of the rod \cite{Brenner1962, Lisicki2016b}, the separation is much larger than the thickness of the rod but much smaller than the length \cite{Katz1975} and the separations is of order of the thickness \cite{Cardinaels2015, JEFFREY1981}. Furthermore each of these solutions were found using different asymptotic techniques; Brenner used the method of reflections to determine the drag when the separation is much larger than all lengths of the rod \cite{Brenner1962}, Katz \textit{et al.} represented the body as a line of point forces above a wall \cite{Blake1971} to determine the drag when the separation is larger than the thickness but smaller than the length \cite{Katz1975}, and Jeffrey and Onishi used lubrication arguments to determine the flow when an infinite cylinder is very close to the wall \cite{JEFFREY1981}. Russel and De Mestre later showed that the model of Katz \textit{et al.} could be extended to capture the results of Brenner \cite{DeMestre1973, DeMestre1975} but as yet no asymptotic representation exists that bridges all the separations. However such representations are still greatly desired { for the modelling of microscopic swimmers near surfaces (biological and artificial) \cite{Walker2019, Das2018, Yang2011, Omori2016, Koens2018a, Elgeti2009}, the dynamics of colloids \cite{Lisicki2016b}, microcantilevers by walls \cite{Clarke2006} and the sedimentation of rods \cite{Russel1977a, Tornberg2006, Holm2007, Tiefenbruck1980, Sendner2007, Zhang2014}. }

In this note we find a representation of the local drag per unit length for a slender rod parallel to a wall which is valid for all separations. { Unlike the equivalent representations for a sphere by a wall \cite{Kim2005, Jeffrey1984a}, this representation of the local drag on a rod does not involve infinite summations or issues with convergence in certain limits.} This representation is produced by asymptotically matching Russel and De Mestre's solution for a rod far from a wall to Jeffrey and Onishi's solution for an infinite cylinder by a wall. We use the matching formalism to determine the error on the infinite cylinder solution and therefore show that the error on the local drag increases towards an asymptotic value as the separation between the wall and rod increases. { We compare these results with numerical solutions, before discussing if} the drag coefficients would be suitable to create a resistive-force theory. { Such a resistive force theory could be used to help understand the planar swimming of spermatozoa \cite{Walker2019} or the dynamics of artificial micro-swimmers which have sunk to the bottom surface \cite{Koens2018a, Zhang2019b}. Finally we demonstrate this resistive force theory on Purcell's two-hinged swimmer,} before concluding the note.

\section{Geometry of a rod parallel to a wall}

 \begin{figure}
\centering
\includegraphics[width=0.6\textwidth]{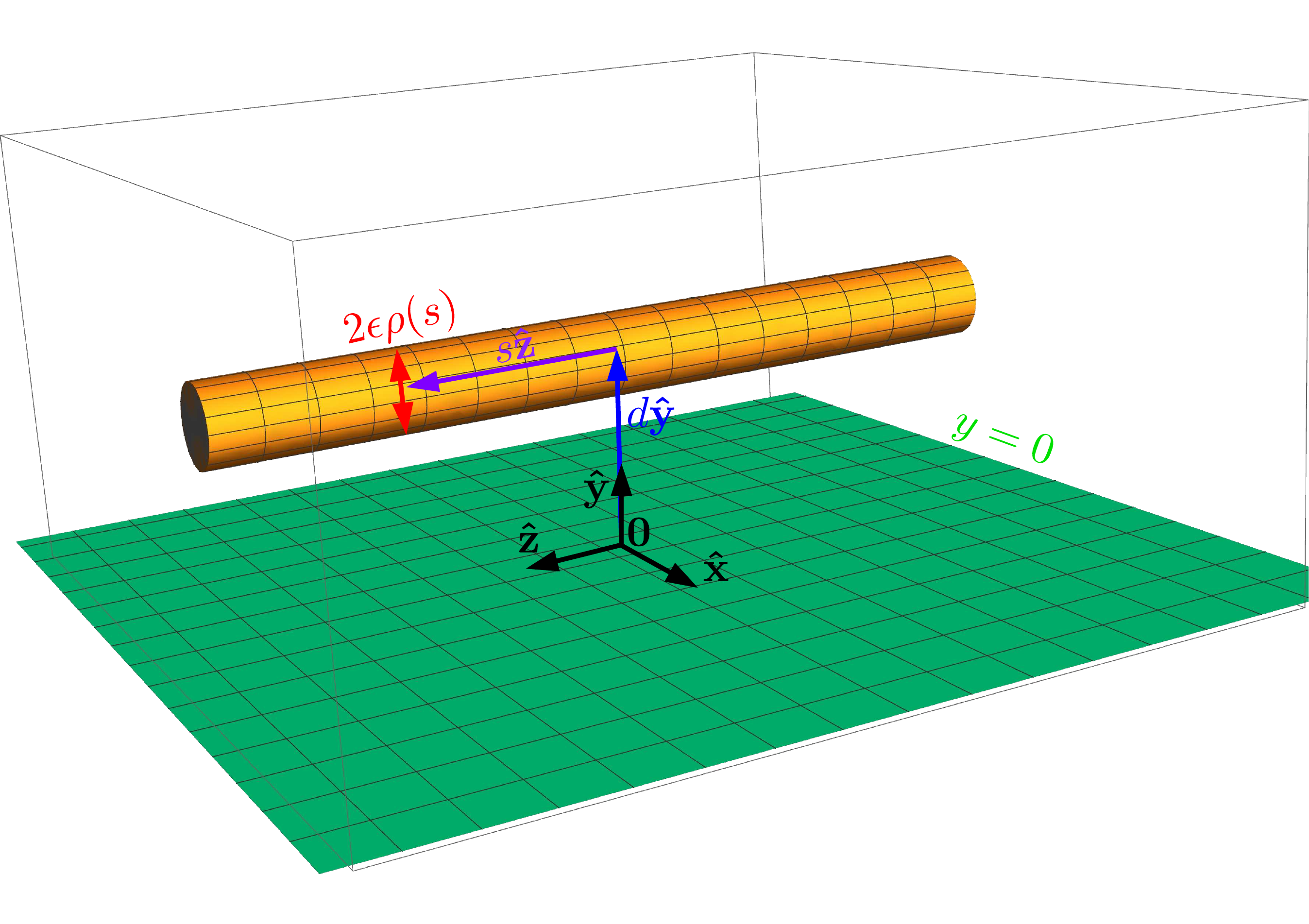}
\caption{Schematic representation of a slender rod above a wall at $y=0$ with $\rho(s) = 1$. The wall is coloured green and the rod is yellow. The origin of the frame is located in the wall while the centre of the rod is a distance $d$ above the origin in the $y$ direction (blue vector). A point along the centreline of the rod is given by  $s \mathbf{\hat{z}}$ (purple) and the thickness of the rod is $2 \epsilon \rho(s)$ (red).} 
\label{fig:diagram}
\end{figure}

In this note we consider a slender cylindrical rod parallel to a plane wall located at $y=0$.  In a Cartesian coordinates system, $\{x,y,z\}$, with the major axis of the rod aligned with the $z$ direction and the centre of the rod located $\{0,d,0\}$ above the wall, the surface of this rod can be parametrised by
\begin{equation}
\mathbf{S}(s,\theta) = \{\epsilon \rho(s) \cos \theta, \epsilon \rho(s) \sin \theta +d, s\}
\end{equation}
where $s \in[-1,1]$ is the arclength, $\epsilon$ is maximal radius of the rod, $\theta$ determines the location on the surface at a given cross-section, and $\epsilon \rho(s) \in [0,\epsilon]$ determines the cross-sectional radius along the length (Fig.~\ref{fig:diagram}).  This parametrisation corresponds to a slender body in the limit $\epsilon \ll 1$ { and a slender rod when $\rho(s)=1$. The equations derived apply in this rod limit, away from the ends, but we will leave $\rho(s)$ arbitrary as it proves useful for estimations of the error.}

\section{The local drag on a rod parallel to a plane wall}

The drag on a rod parallel to a plane wall critically depends on the separation distance, $d$. This is because the asymptotic flow around a slender rod has two regions of behaviour: an inner region in which the rod looks like an infinite cylinder and an outer region in which the rod appears as a line \cite{1976, Koens2018, Keller1976a}. Hence if the distance from the wall is much larger than the thickness of the rod, the wall only contributes to the outer region of the flow, while if $d$ is a similar order to the thickness the wall contributes to the entire flow. The behaviour in each of these cases must therefore be treated separately and then asymptotically matched to create a solution valid for all separations.

\subsection{Drag on a rod far from the wall}

When the distance from the wall is much greater than the radius of the filament, $d \gg \epsilon$, the wall changes the outer flow around the rod. In the outer region an isolated rod looks like a straight line and the flow behaves as a line of point forces \cite{Batchelor2006, 1976}. Russel and De Mestre extended this representation to include the wall by replacing the point forces with point forces by a wall \cite{Blake1971} and expanding the resultant flow in the small $\epsilon$ and $\epsilon/d$ limits \cite{DeMestre1973, DeMestre1975}. This expansion was adapted from Batchelor's work on rods with arbitrary cross-sections \cite{Batchelor2006} and produced an integral equation for the drag per unit length along the body. This drag per unit length on the rod was then expanded in powers of $1/\ln(2/\epsilon)$ and the leading order behaviour determined. They found that the drag per unit length on the filament to $O[1/\ln^3(2/\epsilon)]$ was given by
\begin{equation}
\mathbf{f}^{f}(s) = - \mu \left[\zeta_{1}^{f} \mathbf{\hat{x}} \mathbf{\hat{x}} +\zeta_{2}^{f} \mathbf{\hat{y}} \mathbf{\hat{y}} +\zeta_{3}^{f} \mathbf{\hat{z}} \mathbf{\hat{z}} \right] \cdot \mathbf{U}(s) \label{outer}
\end{equation}
where 
\begin{eqnarray}
\zeta_{1}^{f'} &=& \frac{8 \pi }{\ln \left[ 4(1-s^2)/\epsilon^2 \rho^2(s)  \right]+1- E_1(s) - 2 E_2(s)}, \label{f1} \\
\zeta_{2}^{f'} &=& \frac{8 \pi   }{\ln \left[4(1-s^2)/\epsilon^2 \rho^2(s)  \right] +1 -E_1(s)- 8 E_2(s) + 2 E_3(s)} , \label{f2}\\ 
\zeta_{3}^{f'} &=& \frac{4 \pi }{ \ln \left[4(1-s^2)/\epsilon^2 \rho^2(s) \right]-1 -  E_1(s) + E_2(s)+ E_3(s)}, \label{f3} \\
E_1(s) &=& \mbox{asinh}\left(\frac{1+s}{2 d}\right) + \mbox{asinh}\left(\frac{1-s}{2 d}\right) \\
E_2(s) &=& \frac{1+s}{ 4\sqrt{(1+s)^2+4d ^2}}  + \frac{1-s}{ 4\sqrt{(1-s)^2+4d^2}}, \\
E_3(s) &=& \frac{(1+s)^3}{ 4[(1+s)^2+4d^2]^{3/2}}  + \frac{(1-s)^3}{4 [(1-s)^2+4d^2]^{3/2}}.
\end{eqnarray}
In the above the superscript $f$ is used to denote the solution when $d \gg \epsilon $, $\mbox{asinh }x$ is the inverse of $\sinh x$ and $\zeta_{1}$, $\zeta_{2}$ and $\zeta_{3}$ are the local drag coefficients in $x$, $y$ and $z$, respectively. We note that the typographic errors in Ref.~\cite{DeMestre1975} (see Ref.~\cite{Brennen1977a}) have been corrected above.  The total force on the body from the fluid is then
\begin{equation}
\mathbf{F}^{f} = \int_{-1}^{1} \mathbf{f}^{f}(s) \,ds,
\end{equation}
however, no closed form exists that captures both the limits $d \to \infty$ and $d \to \epsilon$  \cite{Trahan1985}. These results limit to the near drag coefficients derived by Katz \textit{et al.} \cite{Katz1975} as $d \to \epsilon$ and far drag coefficients of Brenner \cite{Brenner1962} when $d \gg 1$. Experiments have also shown that the drag on a rod approaching a plane wall is accurately described by $\zeta_{2}^{f}$ for $d > 3 \epsilon$ \cite{Trahan1985}. This demonstrates the effectiveness of the representation over all $d \gg \epsilon$.

{ 
Though these coefficients are very effective, the dominator of $\zeta_{2}^{f'}$ goes to zero for $d \sim e \epsilon /2$. Hence $\zeta_{2}^{f'}$ only applies if $d \gtrsim e \epsilon /2$ while the coefficients $\zeta_{1}^{f'}$ and $\zeta_{3}^{f'}$ are well-behaved for all $d > \epsilon$. This is mostly likely a reflection of the stronger singularity experienced for the rod approaching the wall rather than moving perpendicular to it. It is possible to overcome this issue by adding a function to the dominator which is smaller than the expansion order of the term. Though only necessary for $\zeta_{2}^{f'}$, it will become apparent that these small functions will help us simplify the matched form and so we include them in all three far coefficients as
\begin{eqnarray}
\zeta_{1}^{f} &=& \frac{8 \pi }{\ln \left[ 4(1-s^2)/\epsilon^2 \rho^2(s)  \right]+1- E_1(s) - 2 E_2(s)+ 2 g_1[\epsilon \rho(s)/d ]},  \\
\zeta_{2}^{f} &=& \frac{8 \pi   }{\ln \left[4(1-s^2)/\epsilon^2 \rho^2(s)  \right] +1 -E_1(s)- 8 E_2(s) + 2 E_3(s) +2 g_2[\epsilon \rho(s)/d ]} , \\ 
\zeta_{3}^{f} &=& \frac{4 \pi }{ \ln \left[4(1-s^2)/\epsilon^2 \rho^2(s) \right]-1 -  E_1(s) + E_2(s)+ E_3(s) + 2 g_3[\epsilon \rho(s)/d ]}, \\
\end{eqnarray}
where the $g_i[\epsilon \rho(s)/d ]$ are small functions to be determined through matching the solutions. We assume that $g_i[\epsilon \rho(s)/d ] = O(\epsilon /d)$ or smaller  and so its contribution to the total drag coefficient is at most $O(\epsilon /d)$, which in the limit $d \gg \epsilon$  is much smaller than the accuracy of the expansion, $O[1/\ln^3(2/\epsilon)]$. Hence the addition of these functions do not significantly change the order of the expansion or the accuracy of the results in its region. The accuracy of this assumption will be checked in the matching.
}

\subsection{Drag on rods near a wall}

When the distance from the wall is similar to the radius of the rod, $d \sim \epsilon$, the wall influences the entire flow. In this limit $x$ and $y$ scales with $\epsilon$, and $z$ scales with $1$.  As a result, the Stokes equations can be asymptotically expanded in terms of a power series in $\epsilon$. This expansion separates the flow and derivatives along the rod's axis, $w$ and $\partial_z$, from the flow and derivatives perpendicular to it, $\mathbf{u}_{\perp}= \{u,v\}$ and $\nabla_{\perp} =\{\partial_x,\partial_y\}$, thereby making the problem two dimensional. Specifically if the flow $\mathbf{u}=\{u,v,w\}$ is expanded as
\begin{eqnarray}
   \mathbf{u} &=& \mathbf{u}^{(0)} + \epsilon   \mathbf{u}^{(1)} + \cdots, \\
p &=& p^{(0)} + \epsilon   p^{(1)} +  \cdots,
 \end{eqnarray} 
 the equations for $\mathbf{u}^{(0)}$ become
 \begin{eqnarray}
\nabla_{\perp} \cdot \mathbf{u}_{\perp}^{(0)} &=& 0, \label{u1}\\
\nabla_{\perp}^{2} \mathbf{u}_{\perp}^{(0)}   - \nabla_{\perp} p^{(0)} &=& \mathbf{0}, \label{u2}\\
\nabla_{\perp}^{2} w^{(0)} &=&0, \label{u3}
\end{eqnarray}
with the boundary conditions
 \begin{equation}
 \mathbf{u}^{(0)}\left(\rho(s) \cos \theta,\rho(s) \sin \theta + d/\epsilon, s \right) = \mathbf{U},  \quad
\mathbf{u}^{(0)}\left(y=0\right) = \mathbf{0},  \quad
\mathbf{u}^{(0)}\left(|\mathbf{x}| \to \infty\right) = \mathbf{0}. 
 \end{equation}
  These equations describe the two dimensional flow around an infinite cylinder of radius $\rho(z=s)$ a distance $d/\epsilon$ above a plane wall. Jeffrey and Onishi \cite{JEFFREY1981} solved these equations for motion perpendicular to $z$, Eqs.~\eqref{u1} and \eqref{u2}, using bipolar coordinates. We omit the full flow for brevity but note that the pressure is
\begin{eqnarray}
p^{(0)}&=& \frac{\mu V }{a(\alpha_1 - \tanh \alpha_1)} \left[2 \cos \beta \cosh \alpha -3 + \frac{2 \cos \beta \cosh(\alpha-\alpha_1) - \cos 2 \beta \cosh (2 \alpha - \alpha_1)}{\cosh \alpha_1} \right] \notag \\
&& + \frac{\mu U }{a \alpha_1 \sinh \alpha_1}\left[ 2 \cosh \alpha \cosh \alpha_1 \sin \beta - \cosh(2 \alpha-\alpha_1) \sin 2 \beta \right],
\end{eqnarray}
where $ \mathbf{U} = \{ U, V, W\}$, $\alpha_1 = \ln \left[ (d + \sqrt{d^2 - \epsilon^2 \rho^2(z)})/\epsilon \rho(z) \right]$ , $\tanh \alpha = 2 a y / (x^2+y^2+a^2)$, $\tan \beta = - 2 a x / (x^2+y^2-a^2)$   and $a^2 = (d/\epsilon)^2  -\rho^2(z)$. 
This pressure will be needed to estimate the error on the leading solution.  Similarly the out of plane motion can also be solved using bipolar coordinates \cite{Yuan2015a} to find
 \begin{equation}
 w^{(0)} = W \frac{\alpha}{\alpha_1},
 \end{equation}
 and the drag per unit length on the body from these flows are
\begin{equation}
\mathbf{f}^{n}(s) = - \mu \left[\zeta_{1}^{n} \mathbf{\hat{x}} \mathbf{\hat{x}} +\zeta_{2}^{n} \mathbf{\hat{y}} \mathbf{\hat{y}} +\zeta_{3}^{n} \mathbf{\hat{z}} \mathbf{\hat{z}} \right] \cdot \mathbf{U}(s) \label{inner}
\end{equation}
where 
\begin{equation}
\zeta_{1}^{n} = \frac{4 \pi  }{\alpha_1}, \quad  \zeta_{2}^{n} = \frac{4 \pi    }{\alpha_1 - \tanh \alpha_1} , \quad \zeta_{3}^{n} = \frac{2 \pi }{\alpha_1},  \label{near}
\end{equation}
$\alpha_1 = \ln \left[ (d + \sqrt{d^2 - \epsilon^2 \rho^2(z)})/\epsilon \rho(z) \right]$, and the superscript $n$ indicates the $d\sim \epsilon$ region. Again the total force is determined by the integration of this result. These coefficients produce the correct lubrication behaviour as $d \to \epsilon$  \cite{Cardinaels2015, JEFFREY1981}. 

The error on these coefficients can be estimated from the $\mathbf{u}^{(1)}$ flow. This flow satisfies
\begin{eqnarray}
\nabla_{\perp} \cdot \mathbf{u}_{\perp}^{(1)} &=& - w_{z}^{(0)},  \label{e1}\\
\nabla_{\perp}^{2} \mathbf{u}_{\perp}^{(1)}  - \nabla_{\perp} p^{(1)} &=& \mathbf{0}, \label{e2}  \\
\nabla_{\perp}^{2} w^{(1)}  &=& p_{z}^{(0)}, \label{e3} 
\end{eqnarray} 
with all the boundaries held stationary. These equations describe the two dimensional viscous flow around an infinite cylinder in the presence of a plane wall with sources of fluid. The force on the body from this flow can be determined using the Lorentz reciprocal relationship \cite{Kim2005}. This relationship relates any two flows that share the same domain and takes the form
\begin{equation} 
 \iint \left(\mathbf{u}\cdot\boldsymbol{\sigma}'\cdot \mathbf{n} -\mathbf{u}'\cdot\boldsymbol{\sigma}\cdot \mathbf{n} \right)\,dS =\iiint\left[ p \nabla\cdot \mathbf{u}'  - p' \nabla\cdot \mathbf{u} +\mathbf{u}\cdot \left( \nabla\cdot \boldsymbol{\sigma}' \right)  - \mathbf{u}'\cdot \left( \nabla\cdot \boldsymbol{\sigma} \right) \right] \,d V 
 \end{equation}
 where the primed and unprimed variables denote different flows with 0 constant background pressure, $\boldsymbol{\sigma}$ is the fluid stress, the surface integrals are taken over all the boundaries and the volume integrals are taken over the entire fluid. If we set, $\mathbf{u}=\mathbf{u}^{(0)}$, and $\mathbf{u}'= \mathbf{u}^{(1)}$, this relationship becomes
 \begin{equation}
 \mathbf{U}\cdot \mathbf{F}^{(1)} = \iiint\left[ w^{(0)}p_z^{(0)}-p^{(0)} w_{z}^{(0)} \right] \,d V + O(\epsilon), \label{lorentz}
 \end{equation}
where $\mathbf{F}^{(1)}$ is the total force on the rod from the correction flow. The boundary condition that the fluid goes to rest as $|\mathbf{x}| \to \infty$ ensures that $p^{(0)}$ and $ w^{(0)}$ decay sufficiently to enable this integral to converge. This integral shows that force on the rod from the $O(\epsilon)$ flow scales with $\partial_{z}(p^{(0)} w^{(0)})\,dV$. Hence the total drag from the next order flow is roughly
\begin{equation}
\epsilon \mathbf{F}^{(1)} = O\left(\frac{d \rho'(z) }{ \rho(z) \alpha_1^{3}} \right) \label{error}
\end{equation}
where we have used  $\partial_z \alpha_1 = - d \rho'(z)/\epsilon \rho(z) a$ and that the pressure, $p^{(0)}$, scales with $1/a \alpha_1$, the axial velocity, $w^{(0)}$, scales with $1/\alpha_1$ and the cross-sectional volume element, $dV$, scales with $a^2$. We note that the total drag at $O(\epsilon)$ also includes contributions from the out of plane components of the normal director on the rods surface. { The above estimate shows that the inner region expansion error scales with $\rho'(z)$ (or any other variation in the geometry with $z$). Hence this error is minimal near the centre of a rod parallel to a wall. However near the ends of any slender body this variation is likely to be very large. Provided these regions of high variation occur over a small fraction of the whole rod, their total contribution to the drag remains small and so the errors can be ignored. 

 This estimate of the error also suggests a non-monotonic dependence on $d$. This is because $\alpha_1 \to \sqrt{d-\epsilon}$ as $d \to \epsilon$ and $\alpha_1 \to \log(d/\epsilon)$ as $d/\epsilon \to  \infty$. Hence as $d/\epsilon$ increases the drag grows almost linearly while as $d \to \epsilon$ the the error grows roughly as $d \rho'(z)(d-\epsilon)^{-3/2}$. This complex behaviour with $d$ reflects the different hydrodynamic behaviours present in the problem. As $d$ increases the initial assumptions underlying this expansion region break down and the total length of the rod becomes significant, hence generating an error with increasing $d$. Conversely, as the body gets close to the wall, the lubrication singularity makes the local geometry critical \cite{ClaeysIBrady1989}. For an infinite cylinder approaching a wall the lubrication singularity goes as $(d-\epsilon)^{-3/2}$ while for a sphere it goes as $(d-\epsilon)^{-1}$. Hence these coefficients can have a large errors close to the wall if the body is not approximately a rod.}

\subsection{The common behaviour}

The different region solutions can be matched together if they share the same behaviour in an overlapping region. This overlap is found by considering the drag coefficients for a rod far from the wall in the limit $d \to \epsilon$ and the drag coefficients for a rod near to a wall in the limit $d/\epsilon \to \infty$. In both these limits the drag coefficients become
\begin{eqnarray}
\zeta_{1}^{c} &=& \frac{4 \pi  }{\ln \left[ 2 d/\epsilon\rho(s)  \right] +g_1[\epsilon \rho(s)/d ]},  \\
\zeta_{2}^{c} &=& \frac{4 \pi   }{\ln \left[2 d/\epsilon \rho(s)  \right] -1 +g_2[\epsilon \rho(s)/d ] }, \\ 
\zeta_{3}^{c} &=& \frac{2 \pi }{ \ln \left[2 d/\epsilon \rho(s) \right]+g_3[\epsilon \rho(s)/d ]},
\end{eqnarray}
{ where the superscript $c$ above denotes the common behaviour and the expansion of the inner solution in the limit $d/\epsilon \to \infty$ reveals
\begin{eqnarray}
g_1\left(\frac{\epsilon \rho(s)}{d} \right) &=&g_3\left(\frac{\epsilon \rho(s)}{d} \right) \notag \\
&=&  \alpha_1 - \ln \left[\frac{2 d}{\epsilon \rho(s)}\right], \\
g_2\left(\frac{\epsilon \rho(s)}{d} \right) &=& \left( \alpha_1 - \tanh \alpha_1\right)- \left(\ln \left[\frac{2 d}{\epsilon \rho(s)} \right] -1\right).
\end{eqnarray}
The relative size of these $g_i[\epsilon \rho(s)/d]$  functions can be determined using Taylor series to find $g_i[\epsilon \rho(s)/d] = O(\epsilon^2/d^2)$ for every $i$. Hence, these functions are smaller than the expansion order in the far from wall region as per our initial assumption. Importantly this is only possible because the leading contributions from Dr Mestre \cite{DeMestre1973, DeMestre1975} and Jeffrey and Onishi \cite{JEFFREY1981} match in these limits. Without this condition we would have found $g_i[\epsilon \rho(s)/d] = O(1)$ which is inconstant with our initial assumptions. The from of these leading contributions can be found if by dropping the $g_i[\epsilon \rho(s)/d]$ functions and are}
the resistance coefficients found by Katz \textit{et al.} \cite{Katz1975}.  Since both these limits are the same, the two solutions match and a composite representation of the drag can be formed \cite{Hinch1991}.
 We note that the matching of the $\zeta_2$ coefficients{, without the $g_i[\epsilon \rho(s)/d]$ functions,}  was previously observed by Trahan and Hussey when they compared the different models to experiments of a rod falling towards a wall \cite{Trahan1985}. 

\subsection{A representation of the local drag which is valid for all separations}

Since the drag coefficients match in the suitable limits, a composite representation for the drag per unit length which is valid for all separations can be created by adding the far and near behaviours together and subtracting off the common behaviour ($\mathbf{f} \approx \mathbf{f}^{f}+\mathbf{f}^{n}-\mathbf{f}^{c}$) \cite{Hinch1991}. Hence the force per unit length on a rod parallel to a plane wall can be asymptotically represented by
\begin{equation}
\mathbf{f}(s) = - \mu\left[\zeta_{1} \mathbf{\hat{x}} \mathbf{\hat{x}} +\zeta_{2} \mathbf{\hat{y}} \mathbf{\hat{y}} +\zeta_{3} \mathbf{\hat{z}} \mathbf{\hat{z}} \right] \cdot \mathbf{U}(s) \label{vd}
\end{equation}
where
{
\begin{eqnarray}
\zeta_{1} &=& \frac{8 \pi  }{\ln \left[ (1-s^2)/d^2  \right]+1- E_1(s) - 2 E_2(s) + 2 \alpha_1}, \label{m1} \\
\zeta_{2} &=& \frac{8 \pi    }{\ln \left[(1-s^2)/d^2 \right] +3 -E_1(s)- 8 E_2(s) + 2 E_3(s)+ 2(\alpha_1 - \tanh \alpha_1)} , \label{m2} \\
\zeta_{3} &=& \frac{4 \pi  }{ \ln \left[(1-s^2)/d^2  \right]-1 -  E_1(s) + E_2(s)+ E_3(s) + 2 \alpha_1} , \label{m3}
\end{eqnarray}
 we have substituted in the $g_{i}[\epsilon \rho(s)/d]$ functions, used $\alpha_1 = \ln \left[ (d + \sqrt{d^2 - \epsilon^2 \rho^2(z)})/\epsilon \rho(z) \right]$ and simplified the equations. From the above we see that through including the $g_{i}[\epsilon \rho(s)/d]$ functions, the near wall and the common behaviour cancel, leaving a single set of coefficients. } The leading order drag on a rod parallel to a plane wall at any separation is therefore 
\begin{equation}
\mathbf{F} =\int_{-1}^{1} \mathbf{f}(s) \,ds.
\end{equation} 
Again no closed form exists for the net drag that bridges all configurations without further approximations \cite{Trahan1985}. { These drag coefficients apply in the limit of a slender rod, $\rho(s)=1$. Hence in order to apply to non-rod like bodies we require $\rho'(z)$ to be small. For simple shapes, like prolate spheroids where $\rho(s) = \sqrt{1-s^2}$, this condition is often met away from the ends of the body. Hence the total drag from a prolate spheroid could be estimated using these coefficients provided the regions of high variation is small relative to the entire length.} Interestingly around $d \sim 1$ the error on the near and far region both scale as $1/ \log^3(2/\epsilon)$. Hence the error on these coefficients appears to increase to $O[1/ \log^3(2/\epsilon)]$ as the separation increases.

{
\subsection{Comparison with established limits and numerical simulations}

The behaviour of the matched resistance coefficients, Eqs.~\eqref{m1}, \eqref{m2} and \eqref{m3}, visibly approaches the well-established limiting behaviour of a rod above a wall in the different respective regions (Fig.~\ref{fig:Forces}). At small separations from the wall and small $\epsilon$, the coefficients are seen to closely replicate the lubrication behaviour on the rod, while for small $\epsilon$ and larger $d$, the far-field behaviour matches better; this behaviour is by construction. However, the plots also reveal that for $\epsilon \sim 0.1$ and $d$ close to the wall, in the matched coefficients there is over 20\% error with both of the established limiting behaviours. This indicates that the model breaks down in this region, and so the matched representation is no longer accurate. Given $\epsilon\sim 0.1$ is typically considered to beyond the accuracy of most local drag models, this behaviour is to be expected.

\begin{figure}
\centering
\includegraphics[width=\textwidth]{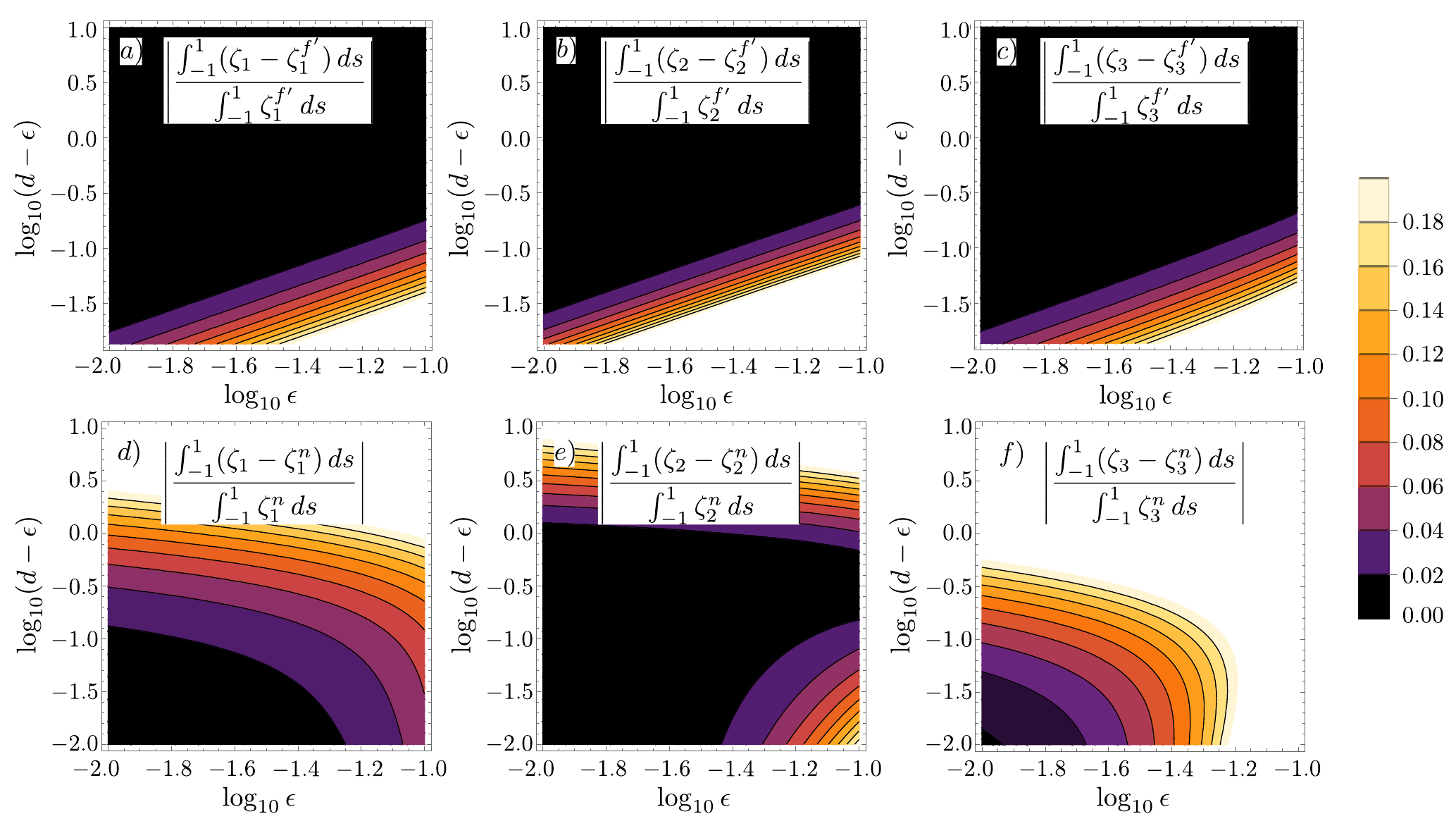}
\caption{Contour plots of the absolute relative difference between the matched drag coefficients and the limiting behaviours for a rod ($\rho(s)=1$). a) $ \left|\int_{-1}^{1} (\zeta_1-\zeta_1^{f'})\,ds /\int_{-1}^{1} \zeta_1^{f'} \,ds \right|$, b)  $\left| \int_{-1}^{1} (\zeta_2-\zeta_2^{f'})\,ds /\int_{-1}^{1} \zeta_2^{f'} \,ds\right|$, c)  $\left| \int_{-1}^{1} (\zeta_3-\zeta_3^{f'})\,ds /\int_{-1}^{1} \zeta_3^{f'} \,ds\right|$, d) $ \left|\int_{-1}^{1} (\zeta_1-\zeta_1^{n})\,ds /\int_{-1}^{1} \zeta_1^{n} \,ds\right|$, e) $ \left|\int_{-1}^{1} (\zeta_2-\zeta_2^{n})\,ds /\int_{-1}^{1} \zeta_2^{n} \,ds\right|$, and f) $\left| \int_{-1}^{1} (\zeta_3-\zeta_3^{n})\,ds /\int_{-1}^{1} \zeta_3^{n} \,ds\right|$.} 
\label{fig:Forces}
\end{figure} 

The accuracy of the matched representation for drag can be further investigated via comparison with numerical simulation of a translating prolate spheroid above a no-slip boundary. Simulations were performed via a constant panel single-layer Boundary Element Method (BEM) with a kernel given by a regularised Blakelet \cite{Smith2009}. Briefly, we begin by constructing a quadratic triangular mesh of a prolate spheroid by projecting a regular mesh of a cuboid, of the same aspect ratio, onto prolate spheroidal coordinates. This produces a high-quality mesh with relatively uniform element areas {(Fig.~\ref{fig:mesh_prp_drag})}. We then specify the velocity of the spheroid $\mathbf{u}(\mathbf{y}_m)$ at the centroids $\mathbf{y}_m$ of each of the $N$ elements $S_{n=1,\ldots,N}$ of the meshed surface $S$. The unknown force per unit area $\mathbf{f}_n$ is approximated as constant over each element, so that our task is to find a solution to the integral equation,
\begin{equation}
	\mathbf{u}(\mathbf{y}_m) = \frac{1}{8\pi\mu}\sum_{n=1}^N{\mathbf{f}_n}\cdot\int_{S_n}\mathbf{B}^\delta(\mathbf{y}_m,\mathbf{x})\,\mathrm{d}S_\mathbf{x}.
\end{equation}
where the regularised Blakelet tensor, originally found by Ainley et al., but containing a minor typographical error \cite{Ainley2008}, is given by Smith \cite{Smith2009}. The integrals of regularised Blakelets over each element are performed via adaptive Fekete quadrature \cite{Montenegro-Johnson2015}, with 10 points for $n \neq m$ and 190 points for the nearly singular integrals $n = m$. Simulations are normalised such that the minor semi-axis of the spheroid is 1, and the regularisation parameter for the Blakelets $\delta = 1/50$. The total drag in any given direction is then given by 
\begin{equation}
 \mathbf{F}=	\sum_{n=1}^N{\mathbf{f}_n}\int_{S_n}\mathrm{d}S_\mathbf{x},
\end{equation}
namely, multiplying the forces by the element areas and summing. This method is accurate, and converges rapidly. {The convergence of our code is examined for a prolate spheroid with $\epsilon = 0.2$ a distance $d = 0.1$ from the boundary (the closest approach modelled) in Fig.~\ref{fig:convergence}. Panel (a) shows the percentage change between subsequent mesh refinements for 16, 32, 48, 64, and 80 elements in the azimuthal direction, corresponding to a range of 832 to 18240 elements in total, with run times ranging from 2 to 840 seconds on a Lenovo Thinkstation with an Intel Xeon W-3265 2.7GHz CPU and 128GB of RAM. The worst errors occur for the perpendicular component of a spheroid being pulled away from the surface, and the percentage change in the value of this drag between 64 and 80 azimuthal elements is 0.16\%, indicating that we have reached convergence.}

{However, such high azimuthal resolution quickly becomes computationally impractical for more slender objects, as doubling the slenderness doubles the number of elements, increasing the memory requirements for storing the dense matrix for the linear system by a factor of 4, and the solution time approximately by a factor of 8. As such, for any given simulation, we take a different approach whereby we run two relatively coarse discretisations, and then perform Richardson extrapolation (used recently by \cite{Gallagher2021} to decrease the regularisation error of the method of regularised stokeslets \cite{Cortez2005}), on the outputs of the two simulations to increase the accuracy of our solution to within an acceptable tolerance of the converged solution. Richardson extrapolation is viable, because we know the order of convergence of our results: in our case, the total drag is a two-dimensional surface integral of the tractions, which is calculated (via the constant panel method) by a 2D mid-point rule, which converges as $h^4$ for element length $h$. Therefore, for all spheroids and distances to the wall we run our simulations for $28$ and $32$ azimuthal elements, producing values $F_z^{\{28\}}$ and $F_z^{\{32\}}$, from which we obtain the new approximation,
\begin{equation}
  F_z^R = \frac{t^4 F_z^{\{32\}} - F_z^{\{28\}}}{t^4 - 1},\quad \text{for} \quad t = 8/7.
\end{equation}
This approximation is within 1.6\% of the converged solution for a prolate spheroid with $\epsilon = 0.2$ at a height of $0.1$ above the boundary, as shown in Fig.~\ref{fig:convergence}b. Though better results can be obtained using $F_z^{\{48\}}$ and $F_z^{\{32\}}$, we deem this sufficiently accurate for the purposes of validating our asymptotic approach.
}
\begin{figure}
    \centering
    \includegraphics[scale = 0.7, viewport=80 130 400 200,clip=true]{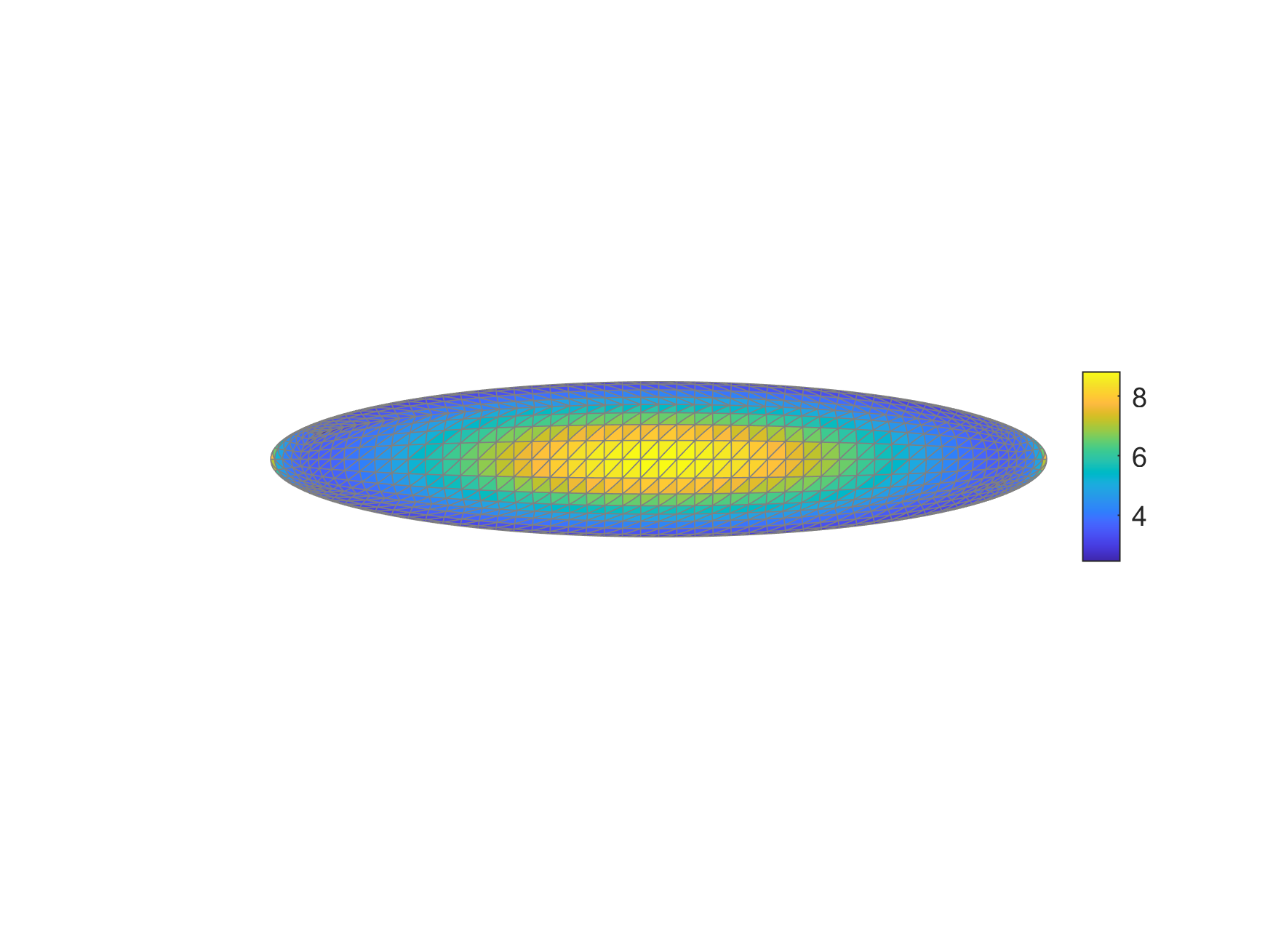}%
    \includegraphics[scale = 0.7, viewport=80 130 400 200,clip=true]{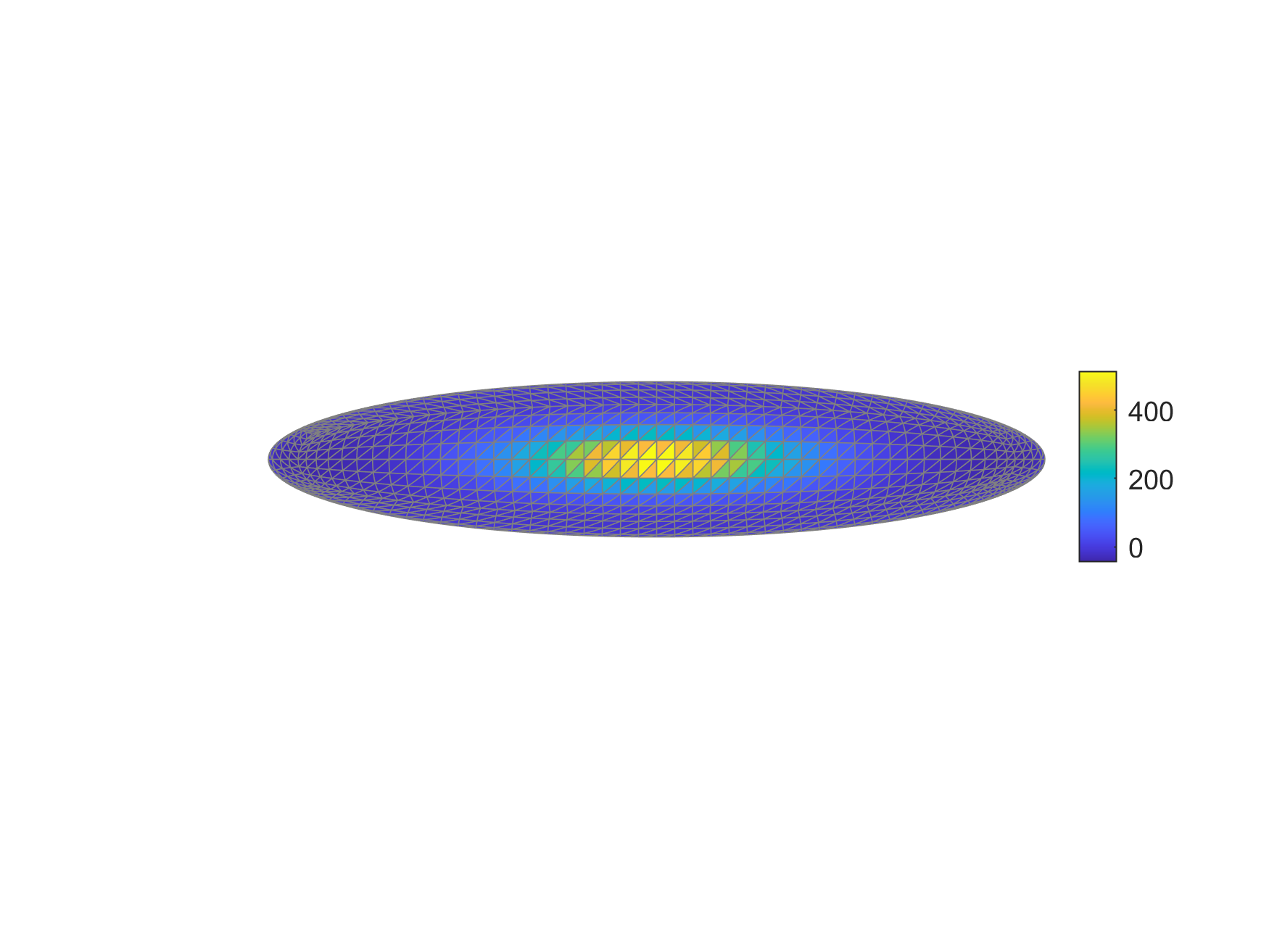}
    \caption{ Perpendicular component of the force per unit area on the underside of a prolate spheroid being pulled away from a wall, in dimensionless units. The azimuthal mesh resolution is 32 elements (3072 elements total), and $\epsilon = 0.2$. The left spheroid is at a height of 1 above the boundary, while the right spheroid is at a height of 0.1, showing the rapid azimuthal variation in the force per unit area for spheroids close to the boundary. The computational mesh is outlined in grey.}
    \label{fig:mesh_prp_drag}
\end{figure}

\begin{figure}
    \centering
    \includegraphics[width=0.95\textwidth]{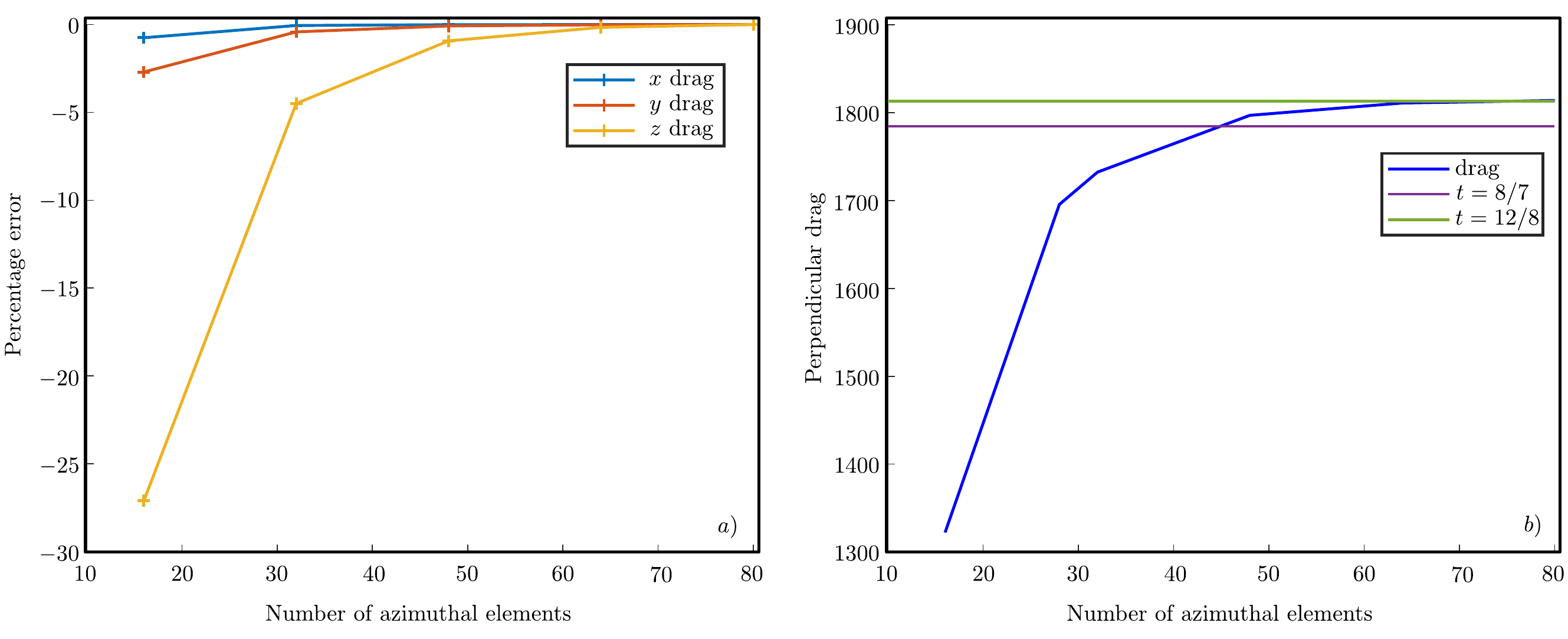}
    \caption{ Convergence of the drag calculations, for a prolate spheroid with $\epsilon = 0.2$, a height of 0.1 above the boundary. a) The percentage difference in the principal drags for 16, 32, 48, and 64 azimuthal elements vs a solution with 80 azimuthal elements, showing good convergence. b) The dimensionless perpendicular component of the drag when the spheroid is pulled away from the wall, showing corrections using Richardson extrapolation for $t = 8/7$ with the solutions for 32 and 28 azimuthal elements being used (resulting in a 1.6\% error), and $t=12/8$ using the more resolved solutions for 48 and 32 azimuthal elements (resulting in a 0.05\% error).}
    \label{fig:convergence}
\end{figure}

Figures~\ref{fig:numerics}(a,b,c) plot the force on the prolate spheroid in each direction as predicted by Eqs.~\eqref{m1}, \eqref{m2} and \eqref{m3} together with the results from the BEM simulations. Visually the model closely replicates the numerical for small $\epsilon$ but starts to differ as $\epsilon$ gets larger. This behaviour is confirmed by the absolute relative error (Fig.~\ref{fig:numerics}d,e,f). Furthermore the error on the model is seen to increase as $d$ decreases. This is because of the lubrication singularities in the drag of a prolate spheroid by a wall differs to that of a rod (which the coefficients model). { This effect is visible by comparing the model to the asymptotic lubrication behaviour of a prolate spheroid (Fig.~\ref{fig:numerics}b) inset black lines) \cite{ClaeysIBrady1989}. In the close region the slope of the model differs to that of the lubrication model, indicating different singularity behaviour. The numerical results, however, agree with the asymptotic lubrication results in this region.} Even with these differences however the relative error is typically 10\% for $\epsilon=0.2$, less than 5\% for $\epsilon=0.1$, and less than 1\% for $\epsilon=0.02$.  This is surprisingly accurate for a local drag representation, as long range hydrodynamic interactions often play an important role to the dynamics of such bodies.

\begin{figure}
\centering
\includegraphics[width=\textwidth]{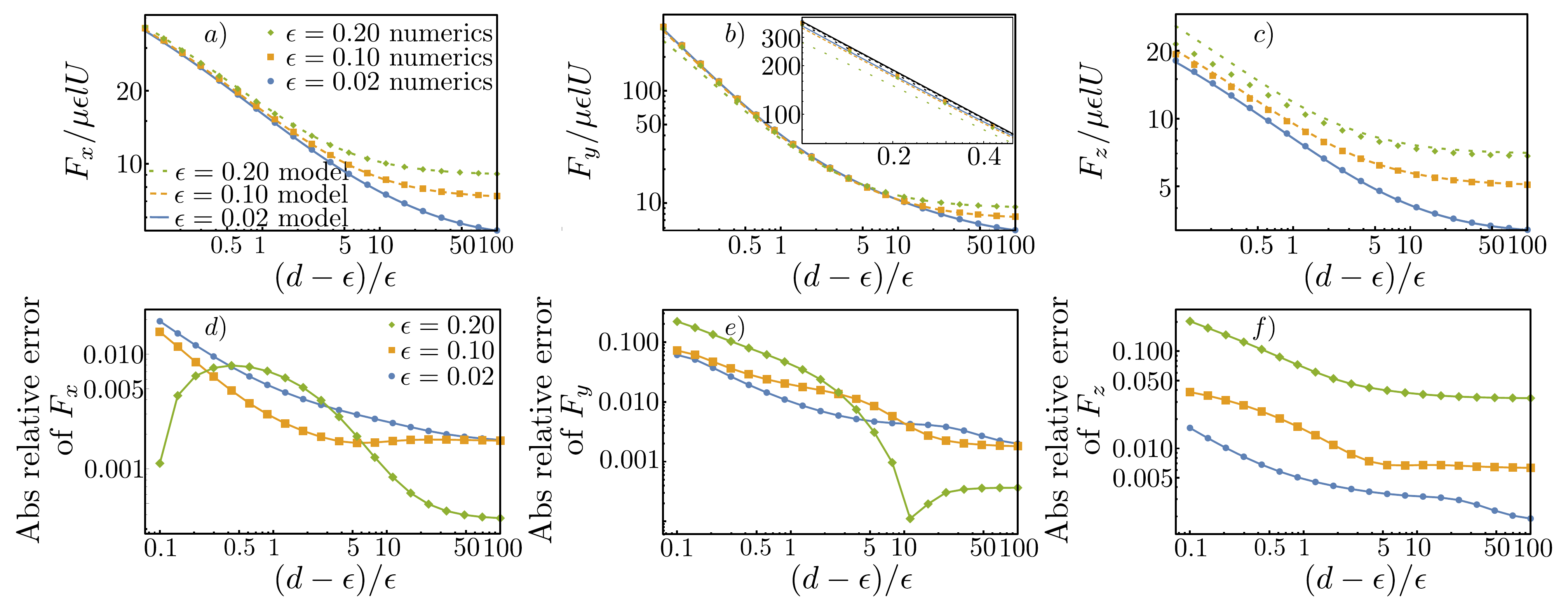}
\caption{The drag on a prolate spheroid parallel to a wall for motion in $\mathbf{\hat{x}}$ (a), $\mathbf{\hat{y}}$ (b), and $\mathbf{\hat{z}}$ (c) and the absolute relative error for each of these direction, (d,e,f) respectively. The absolute relative error is defined as the modulus of the difference between the numerics and the model divided by the numerical result. Blue circles correspond to the numerical results with $\epsilon=0.02$, orange squares correspond to numerical results with $\epsilon=0.1$ and green diamonds correspond to numerical results with $\epsilon=0.2$. In plots (a,b,c) the solid blue line is the model results for $\epsilon=0.02$, the orange dashed line is the model results with $\epsilon=0.1$ and the green dotted line is the model  results for $\epsilon=0.2$. { Black lines in (b) inset represent the asymptotic lubrication behaviour for a prolate spheroid with   $\epsilon=0.02$ (solid), $\epsilon=0.1$ (dashed), and $\epsilon=0.2$ (dotted) \cite{ClaeysIBrady1989}.} } 
\label{fig:numerics}
\end{figure} 

}
\section{The suitability for resistive-force theories}

The above force per unit length representation is suggestive of a resistive-force theory. These theories estimate the drag on a general filament using the drag per unit length along a rod \cite{GRAY1955} and, for filaments in isolation, have errors of $O[1/\ln^2(2/\epsilon)]$ from non-local contributions \cite{Cox}.  { A resistive force theories for filaments perpendicular to walls would be useful for modelling swimmers near boundaries with planar or nearly planer motions \cite{Walker2019, Koens2018a}. These geometries can be surprisingly common as many biological and artificial microswimmers are driven towards walls through hydrodynamics forces \cite{Lauga2009}, the kinematics of their motion \cite{Bianchi2017} or buoyancy effects \cite{Koens2018a, Zhang2019b, Zhang2010}. In this section we discuss the validity of such a resistive force theory and demonstrate its use on Purcell's two-hinged swimmer.
 } 

\subsection{Validity of such a representation}

 The validity of Eq.~\eqref{vd} for general filaments depends on how such non-local contributions change the error in each region. In Russel and De Mestre's model (ie. $d \gg \epsilon$) these non-local factors appear in the integral over the length of the filament \cite{DeMestre1973, DeMestre1975} and, like the isolated case, would produce errors of $O[1/\ln^2(2/\epsilon)]$. Recent comparisons of these coefficients to modern numerical techniques support this \cite{Walker2019}.

The influence of these non-local factors in the near the wall solution is however less obvious. In this case the centreline of the filament can be used to define a set of local coordinates using the the Frenet-Serret formulae \cite{Koens2012a}. Locally these coordinates are cylindrical and so a similar expansion of the Stokes equations in powers of $\epsilon$ produces the same leading order solution. However the error on this solution changes due to the additional sources of variation along the axis of the filament. In this case the Lorentz reciprocal relationship for the force from the correction flow,  Eq.~\eqref{lorentz}, will be modified to account for all the different variations along the length of the filament. Hence provided that these variations are much smaller than $1$, the near wall leading order solution will remain valid. These estimates indicate that the matched drag coefficients can be used for a resistive-force theory for filaments in a plane parallel to a wall, provided there is a slow variation along the length of the slender body. However these coefficients cannot be used for general filaments by a wall because the local geometry of the filament changes with the orientation \cite{Walker2019, Lisicki2016b}.

{
\subsection{Demonstration with Purcell's two-hinged swimmer above a wall}

 \begin{figure}
\centering
\includegraphics[width=\textwidth]{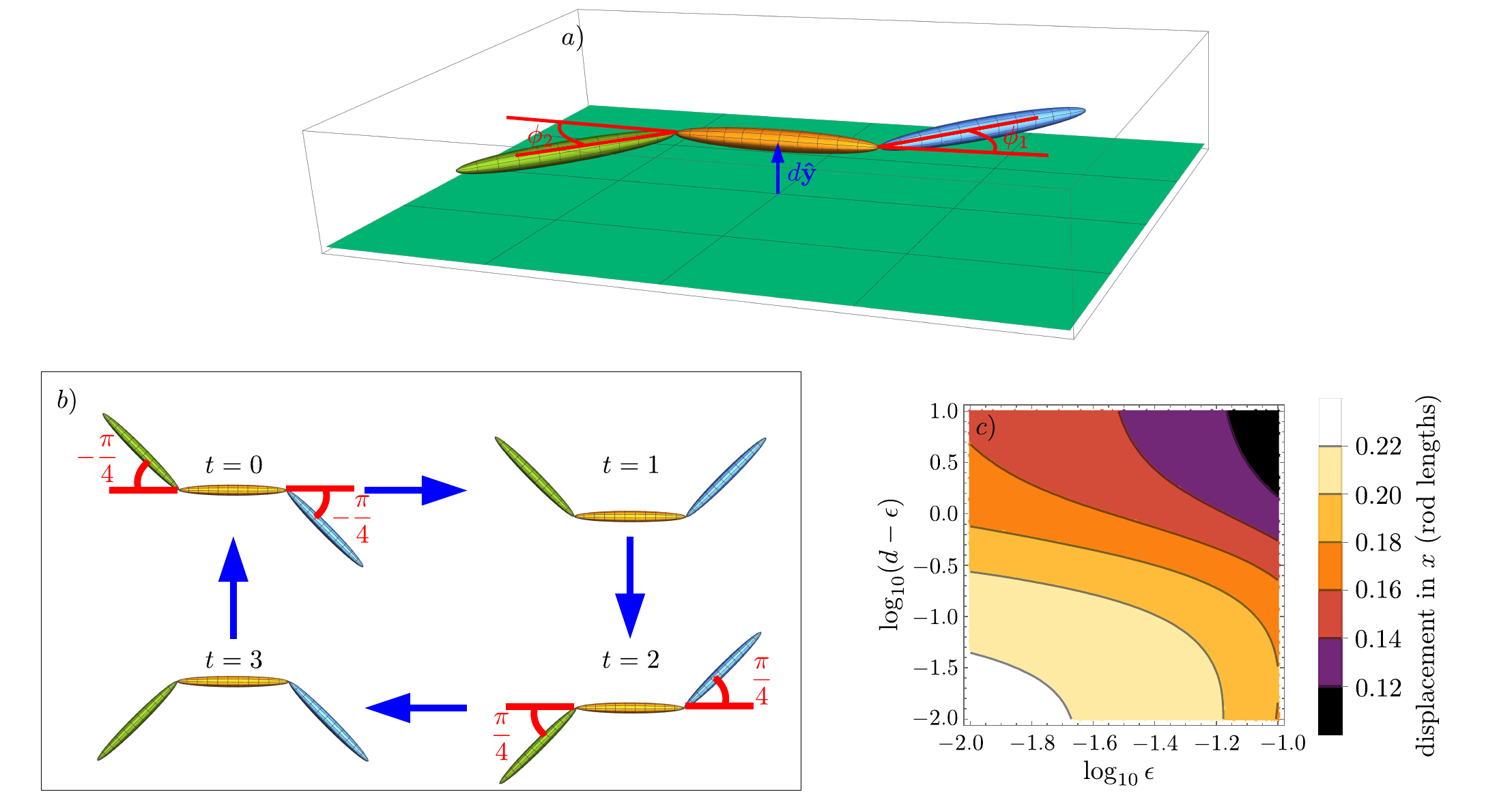}
\caption{a) Schematic representation of a slender Purcell two-hinged swimmer in a plane above a wall with $\rho(s)= \sqrt{1-s^2}$. { Graphical dipiction of the swimming stroke described by Eq.~\eqref{stroke}.  c)} Contour plot of the $x$ displacement of the Purcell swimmer in rod lengths with stroke Eq.~\eqref{stroke} for varying thickness, $\epsilon$, and distance from the wall $d$.} 
\label{fig:purcell}
\end{figure} 

The motion of a Purcell two-hinged swimmer moving in a plane above a wall is one case which could be modelled using these resistance coefficients. This swimmer consists of three rods placed end to end and the angles between the rods can be varied to generate a stroke \cite{Purcell} (Fig.~\ref{fig:purcell}a). This swimmer is often considered a simple prototypical swimmer and its motion has been theoretically studied extensively in an unbounded fluid region using free space resistive force theories \cite{BECKER2003,Gutman2016, Wiezel2018, Avron2008, Hatton2015, Ramasamy2019, Hatton2011, Hatton2013}. However in the experimental realisations of this swimmer there are interfaces and walls are typically present \cite{Chan2009, Kumar2011}. It is therefore interesting to ask how the presence of walls affect the swimmers motion.

The earlier determined drag coefficients, Eq.~\eqref{vd}, allows us to consider a Purcell swimmer moving in a plane parallel to a single flat wall at all separations (Fig.~\ref{fig:purcell}a). Assuming the swimmer is a distance $d$ above a plane wall at $y=0$, the shape of this swimmer can be parametrised by
\begin{eqnarray}
\mathbf{r}_1 &=& \frac{1}{3} \{-3- (2+3s)\cos \phi_1(t) - \cos \phi_2(t), 3d,-(2+3s)\sin \phi_1(t) - \sin \phi_2(t) \}, \\
\mathbf{r}_2 &=&  \frac{1}{3} \{3 s +\cos \phi_1(t) - \cos \phi_2(t), 3d, \sin \phi_1(t) - \sin \phi_2(t) \}, \\
\mathbf{r}_3 &=& \frac{1}{3} \{3+\cos \phi_1(t) +(2+3s) \cos \phi_2(t), 3d,\sin \phi_1(t) +(2+3 s) \sin \phi_2(t) \} ,
\end{eqnarray}
where $s$ is the arclength of each rod, $\mathbf{r}_i$ is the centreline parametrisation of rod $i$, $\phi_1(t)$ and $\phi_2(t)$ are the angles between the rods at time $t$ and we have used the Cartesian coordinates $\{x,y,z\}$.  In the above we have assumed that the rods are thin and so the swimmers shape can described by their centrelines. The surface velocity of each rod can therefore be approximated by $\mathbf{V}_{i} = \partial_t \mathbf{r}_{i} + \mathbf{U} + \boldsymbol{\Omega} \times \mathbf{r}_{i}$, where $\mathbf{U}$  and $\boldsymbol{\Omega}=\omega \mathbf{\hat{y}}$ are the rigid body linear and angular velocities velocity. The resistive force formalism then says the force per unit length along each of these arms is given by
\begin{equation}
\mathbf{f}_i = - \mu \left[\zeta_{1} \mathbf{\hat{n}}_i \mathbf{\hat{n}}_i +\zeta_{2} \mathbf{\hat{y}} \mathbf{\hat{y}} +\zeta_{3} \mathbf{\hat{t}}_i \mathbf{\hat{t}}_i \right] \cdot \mathbf{V}_{i},
\end{equation}
where the index $i$ goes from 1-3, $\mathbf{\hat{t}}_i = \partial_s \mathbf{r}_i$, $\mathbf{\hat{n}}_i = \mathbf{\hat{y}} \times \mathbf{\hat{t}}_i$. Assuming the swimmer is force and torque free the rigid body velocities, $\mathbf{U}$ and $\omega$, can be found by solving
\begin{eqnarray}
\int_{-1}^{1} \left(\mathbf{f}_1+\mathbf{f}_2+\mathbf{f}_3 \right)\,ds &=& \mathbf{0}, \\
\int_{-1}^{1} \left(\mathbf{r}_1\times\mathbf{f}_1+\mathbf{r}_2\times\mathbf{f}_2+\mathbf{r}_3\times\mathbf{f}_3 \right)\,ds &=& \mathbf{0},
\end{eqnarray}
and the trajectory of the swimmer in the laboratory frame is given by
\begin{eqnarray}
\frac{d \mathbf{x}}{d t} &=&\left(\begin{array}{c c c}
\cos \theta & \sin \theta & 0 \\
-\sin \theta & \cos \theta & 0 \\
0 & 0 & 1
\end{array} \right) \cdot \mathbf{U}\\
\frac{d \theta}{d t} &=& \omega
\end{eqnarray}
where $\theta$ is the angle between the laboratory frame $x$ axis and the central rod, $\mathbf{x}(t) =\{X(t),Y(t),Z(t)\}$ is the laboratory frame position. We solve these equations in Mathematica \cite{Mathematica} using a swimming stroke of
\begin{equation}
\{\phi_1(t), \phi_2(t)\} = \left\{ \begin{array}{c r}
\frac{\pi}{4}\{ 2 t -1, -1\} &   0<t<1, \\
\frac{\pi}{4}\{1,  2 (t-1)-1\} &   1<t<2, \\
\frac{\pi}{4}\{ 1 -2  (t-2), 1\} &   2<t<3, \\
\frac{\pi}{4}\{ - 1, 1 - 2 (t-3)\} &   3<t<4,\\
\end{array} \right. \label{stroke}
\end{equation} 
for varying $\epsilon$ and $d$ (Fig.~\ref{fig:purcell}c) with $\rho(s) = \sqrt{1-s^2}$. { This is the classic Purcell swimming stroke stoke (graphically shown Fig.~\ref{fig:purcell}b) and only creates net motion in the $\mathbf{\hat{x}}$ direction due to symmetry \cite{BECKER2003,Purcell}. In the above stoke $t$ is scaled such that $\partial_{t} \phi_{i}=\pi/2$. The choice of this parametrisation does not effect the net displacement and only rescales the swim velocity due to the time-independence and linearity of the Stokes equations \cite{Koens2021}.} Figure~\ref{fig:purcell}c shows that as the Purcell swimmer gets thinner and closer to the wall its displacement per stroke increases. This increase is due to the decreasing thickness and wall separation both increase the drag anisotropy and so promotes larger displacements \cite{BECKER2003,Koens2016a}.
}
\section{Conclusion}

{ The viscous hydrodynamics of bodies near walls is hard to determine but critical to many physical systems. Even in the case of a sphere above the wall a simple representation for the drag which is valid for all separations currently eludes us \cite{Kim2005, Jeffrey1984a}.}
In this note we identified the leading order drag coefficients for a slender rod parallel to a plane wall. Unlike previous models, this representation is valid for all separations above the wall and was found by asymptotically matching the behaviour of rods far from and near to the wall. The results of Russel and De Mestre \cite{DeMestre1973, DeMestre1975} were used for the drag on a slender rod far above a plane wall and the two dimensional flow solution of Jeffrey and Onishi  \cite{JEFFREY1981} was used for the leading order drag on a rod near a wall. We show the error on the near wall solution to increase roughly linearly with the separation from the wall. We then showed that these leading order far and near solutions match in the appropriate limit and so can be combined to produce a composite representation for the drag per unit length along the rod which is valid for all separations. 

{  We then used these coefficients to form a resistive-force theory of a filament perpendicular to a wall. This resistive force theory could be used to better understand the motion of biological and artificial microswimmers near walls which use nearly planar swimming strokes. We demonstrate this with Purcell's two-hinged swimmer and analysed how the swimmers speed changed with its distance from the wall.}

Though this model can handle arbitrary separations from the wall, it restricts the geometry and requires the filament to be exponentially thin when far from the wall. Effective models for slender bodies by a plane wall need to release both these restrictions to be applicable to a large range of problems and is the subject of ongoing work.

\acknowledgements 
LK has received funding from the Australian Research Council (ARC) under the Discovery Early Career Research Award scheme (grant agreement DE200100168) and Macquarie University's new staff grant. T.D.M.-J. gratefully acknowledges support from a Leverhulme Trust Research Leadership Award. The authors also thank Maciej Lisicki and Geordie McBain for useful discussions and advice.

\appendix

\bibliographystyle{ieeetr}
\bibliography{library}

\end{document}